\documentclass[twocolumn,showpacs,preprintnumbers,amsmath,amssymb]{revtex4}

\usepackage{graphicx}
\usepackage{dcolumn}
\usepackage{bm}

\begin{document}


\title{Multiscale reconstruction of time series}

\author{A. P. Nawroth}
\author{J. Peinke}%
\affiliation{%
Institut f\"ur Physik, Carl-von-Ossietzky Universit\"at Oldenburg, D-26111 Oldenburg, Germany
}%

\date{\today}

\begin{abstract}
A new method is proposed which allows a reconstruction of time series based on higher order multiscale statistics given by a hierarchical process. This method is able to model the time series not only on a specific scale but for a range of scales. It is possible to generate complete new time series, or to model the next steps for a given sequence of data. The method itself is based on the joint probability density which can be extracted directly from given data, thus no estimation of parameters is necessary. The results of this approach are shown for a real world dataset, namely for turbulence. The unconditional and conditional probability densities of the original and reconstructed time series are compared and the ability to reproduce both is demonstrated. Therefore in the case of Markov properties the method proposed here is able to generate artificial time series with correct n-point statistics.
\end{abstract}

\maketitle

\section{\label{lab-sectionI} Introduction}

A typical feature of complex systems, like turbulence and finance, is that these have hierarchical and non-trivial structures on different scales. Due to the hierarchical structure such systems are in general very high dimensional, thus there is the challenge to find appropriate methods to model these structures efficiently. Successful attempts have been made to describe these systems as processes in scale rather than time or space itself. Examples are the description of roughness of surfaces \cite{waechter03,jafari2003}, turbulence \cite{friedrich97b,friedrich97,tutkun2004}, earthquakes \cite{tabar2005} and finance \cite{renner01b,ausloos2003}. These successful attempts are characterised by the fact, that they can correctly provide the joint probability density function $p(y_1(\tau_1),...,y_n(\tau_n))$ of the increments $y(\tau)$ of the process variable $x$ at different scales $\tau$,
\begin{eqnarray}
\label{eq_increment}
y(t,\tau):=x(t)-x(t-\tau).
\end{eqnarray}
An increment time series $y(t,\tau)$ for arbitrary scales $\tau$ can be constructed from $x(t)$. In the following the joint statistics of these increment processess are considered. Hereby it is  assumed, that these processess, $y(t,\tau)$, are stationary  in $t$ and the ergodic hypothesis can be applied. Note that the method itself is not restricted to this particular increment definition \cite{karth2003}. For reasons of convenient notation we assume for the following discussion a given complex system $x(t)$, where $t$ denotes time but may denote space as well. The smaller increments are nested into the larger increments.

It should be pointed out, that the knowledge of the joint probability density function $p(y_1(\tau_1),...,y_n(\tau_n))$ includes also information like multifractal scaling properties, typical expressed by $< y^n (\tau) > \sim \tau^{\xi_n}$.

For complex systems there is a general challenge, to generate surrogates based on the knowledge of the statistical properties of the system itself. One approach is to propose general models and proper parameter sets, see for example \cite{breymann2000,muzy2000}. Typically remarkable features like correlations or the shape of single scale pdfs  $p(y(\tau))$ are used for the parameter selection. Some of these approaches provide also a remarkably good description of higher order statistics. Here we propose an alternative, more general approach, which is based on the knowledge of multiscale properties. In particular we show how a given time series can be extended into the future properly.

\section{\label{lab-sectionM} Method}

As stated above a hierarchical complex system can often be described  by means of the joint probability density $p(y_1(\tau_1),...,y_n(\tau_n))$. Because of the involved scales the dimension of the joint probability density is very high. Therefore it is in general very difficult to compute it from empirical time series. However the description and computation can be highly simplified if Markov properties can be assumed. This is the  case if 
\begin{eqnarray}
\label{eq_markov_properites}
p(y_i(\tau_i)|y_{i+1}(\tau_{i+1}),...,y_n(\tau_n)) = p(y_i(\tau_i)|y_{i+1}(\tau_{i+1}))
\end{eqnarray}
is true for all $i$ and $n > i$. Without loss of generality  we take $\tau_{i} < \tau_{i+1}$. It should be noted that the Markov property can be tested for a given data set \cite{renner01b,renner2001,friedrich98}. In this case the joint probability density can be substantially simplified:
\begin{eqnarray}
\label{eq_probability_factor}
\lefteqn{p(y_1(\tau_1),...,y_n(\tau_n)) =  } \hspace{-0.0cm} \\
\nonumber \\
& & p(y_1(\tau_1)|y_2(\tau_2))\cdot ...\cdot p(y_{n-1}(\tau_{n-1})|y_n(\tau_n))\cdot p(y_n(\tau_n)) \nonumber 
\end{eqnarray}

In the following we restrict the discussion to Markov processes in scale and to right-justified increments, i.e. the smaller increment is nested into the larger one and has the right end point in common \cite{waechter2004b} according  to Eq. (\ref{eq_increment}). There are two possible ways to determine the conditional probability densities $p(y_{i-1}(\tau_{i-1})|y_i(\tau_i))$ and the unconditional probability density $p(y_n(\tau_n))$. The first one is straightforward. A typical time series from the complex system is examined, and the probability densities are determined, by simply calculating histograms from the data. Due to the finite length of the data set, very small probabilities tend to be estimated as zero. In order to obtain the joint probability, the conditional and unconditional probability function have to be multiplied. Therefore the joint probability will also be zero for this value.  Due to that, the probability density can be underestimated, especially in case of large negative and positive increments. 

A second possibility to estimate these probabilities is to use the ansatz presented in \cite{renner01b,renner2001,kleinhans2005,mourik2005}. The probability densities are now obtained as a solution of the Fokker-Planck equation:
\begin{eqnarray}
\label{eq_fokker_planck}
\frac{\partial p(y,\tau)}{\partial \tau} = \left [ -\frac{\partial}{\partial y} D^{(1)}(y,\tau)+\frac{\partial^2}{\partial y^2} D^{(2)}(y,\tau) \right ] p(y,\tau).
\end{eqnarray}
Note that this equation holds also for the conditional probability densities.
The Kramers-Moyal coefficients $D^{(i)}$ are defined as 
\begin{eqnarray}
\label{eq_kramers_moyal}
\lefteqn{D^{(i)}(y,\tau) =  } \hspace{-0.0cm} \\
\nonumber \\
& & lim_{\Delta\tau \rightarrow 0} \; \frac{1}{i!\Delta\tau} \int (y'-y)^i \; p(y'(\tau+\Delta\tau)|y(\tau)) \; dy' \nonumber 
\end{eqnarray}
and can again be directly computed from given data. The problem of zero values vanishes, or at least is shifted to increments with a very large magnitude (see for example \cite{renner01b}).

Next we focus on the procedure to create time series. We start with the most general case, that the time under consideration for the new element of the series $(x(t))$ is named $t^*$, then the knowledge of all points $\{x(t^*-\tau_1),...,x(t^*-\tau_n)\}$ and the corresponding probability densities for $\{y(\tau_1),...,y(\tau_n)\}$ are needed in order to choose the value of $x(t^*)$ correctly. Let's assume that $\tilde{x}(t^*)$ is chosen as a value for $x(t^*)$. The corresponding increments $\tilde{y}_i(\tau_i)$, with their common right endpoint $\tilde{x}(t^*)$, are defined through
\begin{eqnarray}
\label{eq_inc_estimate}
\tilde{y}_i(\tau_i):=\tilde{x}(t^*)-x(t^*-\tau_i).
\end{eqnarray}
Because $P(\tilde{y}_n(\tau_n))$ and $x(t^*-\tau_n)$ are known, the probability $P(\tilde{x}(t^*)|x(t^*-\tau_n))$ of this event is also known. But this conditional probability includes only the information of one scale, namely $\tau_n$. To add the  information available on the scale $\tau_{n-1}$,  $p(y_{n-1}(\tau_{n-1})|y_n(\tau_n))$ is also needed. These quantities then determine the probability  $P(\tilde{y}_{n-1}(\tau_{n-1}),\tilde{y}_n(\tau_n))$ and contain the additional information of $x(t^*-\tau_{n-1})$. Therefore now $P(\tilde{x}(t^*)|x(t^*-\tau_n),x(t^*-\tau_{n-1}))$ is known. Due to the Markov property this can easily be extended to the probability $P(\tilde{x}(t^*)|x(t^*-\tau_n),...,x(t^*-\tau_1))$.

Thus the conditional probability density conditioned on many different scales is known. For a finite range of values for $\tilde{x}(t^*)$ and a finite number of necessary scales the probability density $p(x(t^*)|x(t^*-\tau_n),...,x(t^*-\tau_1))$ is obtained. $p(x(t^*)|x(t^*-\tau_n),...,x(t^*-\tau_1))$ now contains all relevant statistical information of the previous time series for a correct choice of the value $x(t^*)$. Choosing now a random value from this distribution, the time series will be extended correctly by another point. Repeating this procedure will produce a time series, which exhibits the correct joint probability density function for all considered scales $\tau_1,...,\tau_n$. Obviously the estimation of $p(x(t^*)|x(t^*-\tau_n),...,x(t^*-\tau_1))$ has to be repeated for each new value of the time series. 
\begin{figure}[t!]
\includegraphics[width= 8.5cm]{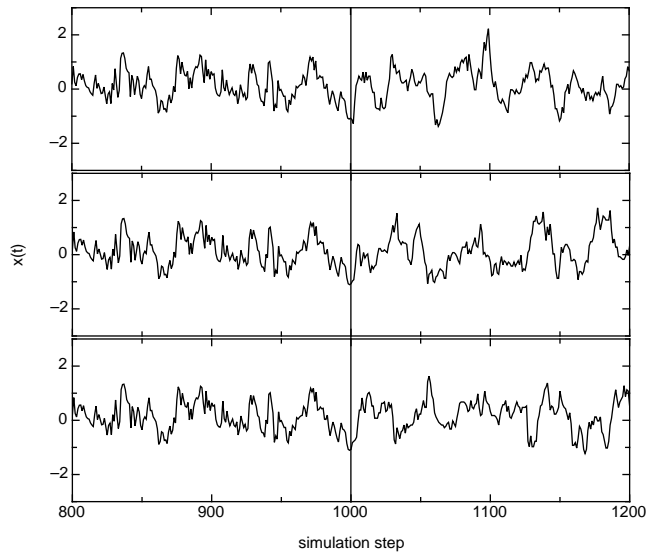}
\caption{\label{fig_000evolution} Three runs of reconstructed time series, which have the same initial conditions. Left to the vertical line, the 200 data points, which were taken as initial condition, are shown. On the right side the following 200 points from our reconstruction are shown.  x(t) has been normalised to zero mean and is presented in units of the standard deviation $\sigma_{\infty}$ of the whole dataset.}
\end{figure}
Some open questions remain: 1. How many and which different scales should be included? 2. How to cope with unavoidable discreteness of the statistics? 3. Which initial conditions should be used for the simulation? The first question of correct choice of scales gets simpler, if for the complex system there exists a large scale, L, above which neighboring increments become statistical independent. A further simplification is given, if there is a small scale, the so called Markov-Einstein coherence length, $l_{mar}$ \cite{lueck2006}, below which the Markov properties are no longer fulfilled. For this case we suggest as number $N$ of scales:
\begin{eqnarray}
\label{eq_inc_estimate}
N=\frac{\log\frac{L}{l_{mar}}}{\log 2}+1.
\end{eqnarray}
In some cases, L can be very large, which is especially true for bounded time series. Here we propose to restrict to the small and medium size increments. The  influence of the large size increments can be included by a specific condition on the  largest  conditional increment. We close the sequence of conditional pdfs as follows: $p(y_1(\tau_1)|y_2(\tau_2))\cdot ...\cdot p(y_{n-1}(\tau_{n-1})|y_n(\tau_n))\cdot p(y_n(\tau_n)|x(t^*-\tau_n))$

\begin{figure*}[t]
\includegraphics[width= 18cm]{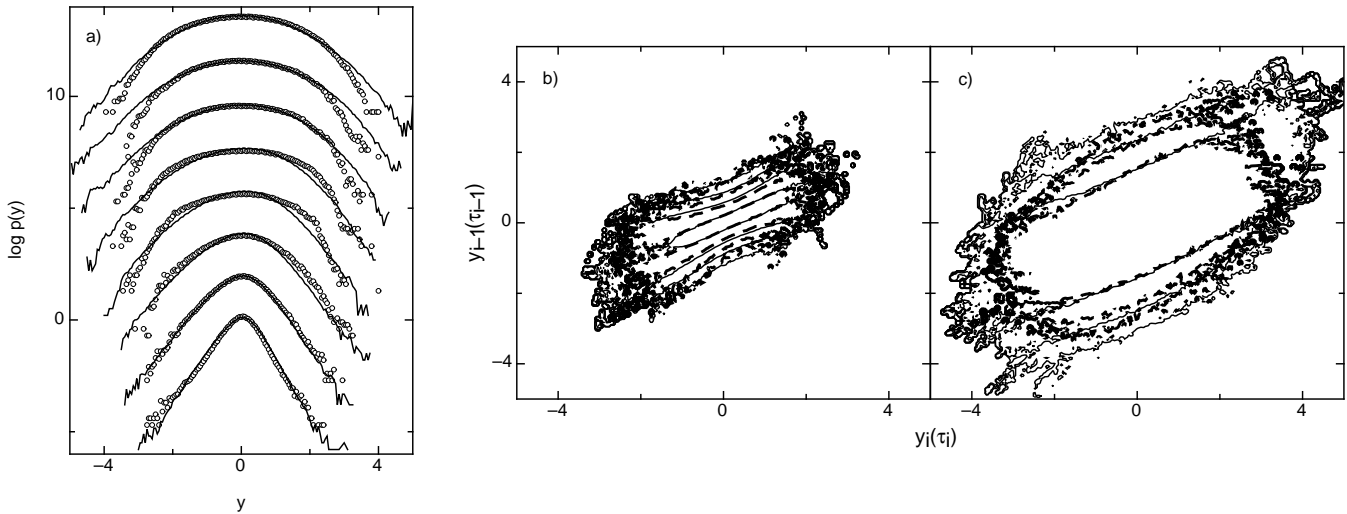}
\caption{\label{fig_000recalc}  a) The pdfs from the original (solid black line) and the reconstructed (circles) data set are shown. The considered scales are  $2^n \cdot  l_{mar}$ with $n=7,6,5,4,3,2,1,0$ from top to bottom.  The pdfs for different scales are shifted for clarity of presentation. b+c) Contour plots of conditional probability densities $p(y_{i-1}(\tau_{i-1})|y_i(\tau_i))$ with $\tau_i=2 \cdot  \tau_{i-1}$ and $\tau_{i-1}=l_{mar}$ (b) and $\tau_{i-1}=2^6 \cdot  l_{mar}$ (c). The ones from original data are denoted as thin solid lines, while the ones from reconstructed data as thick dotted lines. Increments are given in units of the standard deviation $\sigma_{\infty}$ of the whole dataset x(t).}
\end{figure*}

\section{\label{lab-sectionRFT}Results for turbulence}

As a real world example turbulence is chosen, where hierarchical Markov processes were well-demonstrated \cite{friedrich97b,renner2001}. The dataset was produced by a free jet of air into air. The dataset consists of $12.5 \cdot 10^6$ data points, for more details the reader is refered to \cite{renner2001}. A discretisation of the scale and the velocity is chosen as follows. For the smallest scale we choose the above mentioned Markov-Einstein coherence length, $l_{mar}$, in our case  $6mm$. The selected larger scales are $2^n \cdot l_{mar}$. So a reconstruction of the hierarchical process is attempted for the scale interval $l_{mar}$ to $128 \cdot  l_{mar}$, where the largest scale is well above the so called turbulent integral length. The influence of the larger scales is included by the approximation mentioned above. For this choice of scales we are sure, that the necessary Markov properties are always fulfilled. Due to the discretisation of the velocity values, the velocity increments on the smallest scale $l_{mar}$ can have 201 possible values. For the reconstruction in this section, all pdfs were calculated directly from the data without the use of Eqs. (\ref{eq_fokker_planck}) and (\ref{eq_kramers_moyal}).

In order to show typical reconstructed time series, 200 data points from the dataset described above were taken as an initial condition for  the calculation of 200 synthetic data points. The results for three runs are shown in Fig. \ref{fig_000evolution}. Clearly, the three different runs yield similiar results for the first synthetic data points, but become more and more different for the further data points. However the general characteristics of the initial time series are retained. In order to show, how well the multiscale behaviour is captured, a reconstructed time series which contains $10^6$ data points is produced, without using the original data set as initial condition. As an initial condition in this case, a sequence of zeros was used and the first elements of the simulation were discarded. The corresponding probability density functions for the eight different scales are shown in Fig. \ref{fig_000recalc}a and compared to those of the original data.  It is clearly seen, that for $|y| \lessapprox 2 \sigma_{\infty}$ the agreement between the original and reconstructed pdfs is quite good. For extreme values the reconstructed data deviate due to the reasons specified above. With increasing number of given original data the range of agreement in the pdfs extends.

In order to see if the multiscale description is correct, not only the unconditional probability densities have to be examined, but according to Eq. (\ref{eq_probability_factor}) also the conditional probability density of first order. In Fig. \ref{fig_000recalc}b and Fig. \ref{fig_000recalc}c the results are shown for the smallest and the largest scales considered here. In both cases again the empirical conditional probabilities are modelled very well.

\begin{figure}[t]
\includegraphics[width= 6cm]{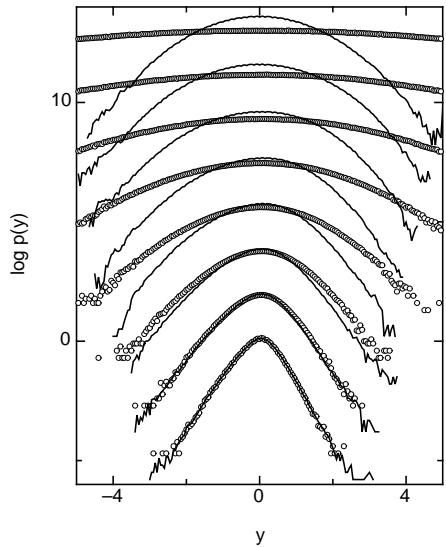}
\caption{\label{fig_ipdf_recalcPerm} Pdfs in analogy to Fig. \ref{fig_000recalc}a  are shown. For the pdfs (circles) of the reconstructed data now a simple one scale process was used. Further parameters are chosen as in Fig. \ref{fig_000recalc}a.}
\end{figure}

In a next step we compare our method with a one scale method to reconstruct time series. Therefore a random walk is constructed, such that the distribution of the increments on the smallest scale is identical to that of the empirical time series, i.e. only $p(y_{1}(\tau_{1}))$ is used. The results for this adapted random walk are shown in Fig. \ref{fig_ipdf_recalcPerm}. As can be seen for small scales the adapted random walk provides good approximations for the increment distribution, but for the larger scales, the approximation with the adapted random walk is very poor. This illustrate the advantage of using multiscale information for the reconstruction. 

\section{\label{lab-sectionC} Conclusions}

A new method has been presented, which allows for a multiscale reconstruction of a time series. More scales are modelled simultaneously, using conditional probabilities of properly chosen increments. A necessary feature of the underlying complex system is that it exhibits Markov properties of the scale dependent process. This can be tested statistically. In this case all conditional probabilities of arbitrary order can be substituted by conditional probabilities of first order, which can be calculated easily from empirical time series. Using the empirical probability densities a time series can be reconstructed. Examples of reconstructed time series for a real world example from turbulence have been shown. The reconstructed time series reproduces the unconditional and the conditional probability densities very well. The advantage of this method is the use of the distribution which are obtained directly from the data. Therefore a modelling of an asymmetric or non-Gaussian distribution is possible. The reconstruction can be applied to any range of scales without a change of the underlying method, as long as Markov properties in scale are present. Furthermore using this approach the full probability density for the next potential step of a time series is provided. Using an empirical time series as a starting point, this procedure can be applied for prediction purposes. But contrary to many other methods not only the mean value for the next step is provided, but the full distribution. This therefore allows also the prediction of volatility or the probability of extreme values. Due to the fact, that only empirical distributions are used, there is no necessity of parameter estimation. Also this method is quite fast, in order to simulate a process with $10^6$ points it takes around one hour on a normal personal computer. In our opinion further improvement can be obtained, if  refined unconditional and conditional pdfs, which provide better approximation of the real distribution for extreme values, are used.

We are very grateful to David Kleinhans, Stephan Barth and Carola von Saldern for interesting discussions and remarks.




\bibliography{AG_LITERATUR}

\end{document}